\documentstyle[aps,amssymb]{revtex}

\begin{document}
\draft
\title{Magnetic and Electric properties of $La_{1-\delta }MnO_{3}$}
\author{S. de Brion$^{1}$, F.\ Ciorcas$^{1}$, G. Chouteau$^{1}$, P. Lejay$^{2}$, P.
Radaelli$^{3}$, C. Chaillout$^{4}$}
\address{$^{1}$Grenoble High Magnetic Field Laboratory\\
MPI-FKF and CNRS \\
BP 166-38042 Grenoble cedex 9, France\\
$^{2}$CRTBT\\
CNRS, BP 166-38042 Grenoble cedex 9, France\\
$^{3}$ILL, BP 156-38042 Grenoble cedex 9, France\\
$^{4}$Laboratoire de Cristallographie\\
CNRS, BP 166-38042 Grenoble cedex 9, France}
\maketitle

\begin{abstract}
The magnetic phase diagram of $La_{1-\delta }MnO_{3}$ powdered samples have
been studied as a function of $\delta $ in the low doping range.\ $%
La_{0.97}MnO_{3}$ has a canted magnetic structure at low temperature $%
(\theta \simeq 130{{}^{\circ }})$. Above $T_{C}=118K$, it becomes a
paramagnet with a huge effective magnetic moment, $\mu _{eff}=6.0\mu _{B}$,
reflecting the presence of magnetoelastic polarons which are not affected by
the magnetic field (up to 20T) nor the temperature $(1.2T_{C}<T<2.5T_{C})$.\
When $\delta $ is increased to $\delta =0.09$, polarons are still present at
high temperature, with a smaller size: $\mu _{eff}=5.8\mu _{B}$. The system
becomes fully ferromagnetic below $170K$ but remains insulating down to the
lowest temperature.
\end{abstract}


\newpage

\section{Introduction}

The manganite compounds $L_{1-x}M_{x}MnO_{3}$ where $L=La,Nd,Pr,...$ and $%
M=Sr,Ba,Ca,...$, exhibit a variety of unexpected properties such as colossal
magnetoresistance\cite{Jin}, field induced structural transition\cite
{Asamitsu (nature)}, charge ordered state\cite{Tomioka (PRL)}, etc... These
properties depend crucially on the doping level as well as the nature of the
doping element\cite{Hwang (95)}. They are closely related to the mixed
valence Mn(III)-Mn(IV).

These compounds have been studied for a long time\cite{Wollan}\cite{Jonker}. 
$LaMnO_{3}$ contains $Mn^{3+}$ ions $\left( S=2\right) $. It is
antiferromagnetic $(T_{N}\simeq 140K)$ and insulating. It becomes
ferromagnetic and metallic upon hole doping (introduction of $Mn^{4+}$ ions)
with the maximum of $T_{C}$ reached when the $Mn^{4+}$ concentration is
around $x\simeq 0.35$.\ C. Zener\cite{Zener (1951)} proposed a mechanism of
double exchange between the itinerant $e_{g}$ electrons and the localized $%
t_{2g}$ electrons forming $S=3/2$ ions $\left( Mn^{4+}\right) $. In this
process, the electrons tend to move from one ionic site to another with the
same spin orientation; therefore electron delocalization favors
ferromagnetism. P.W. Anderson and H. Hasegawa\cite{Anderson (1955)}, P.G.\
de Gennes\cite{de Gennes}, K.Kubo and N.\ Ohata$\ $\cite{Kubo (1972)}
further developed this theory. Recent calculations\cite{Millis (1995)} have
shown that double exchange alone cannot account for the observed colossal
magnetoresistance. Magneto-elastic effects are very important in these
compounds\cite{Asamitsu (nature)}\cite{Bishop1}. Several theoretical
attempts have been made to incorporate electron-phonon coupling{\it \ }\cite
{Millis (1996)}\cite{Roder1996} leading to the idea of lattice polarons
above the ferromagnetic transition{\it . }However, this lattice polaron
picture remains controversial: C.M. Varma \cite{Varma} proposed rather the
idea of spin polarons driven by thermal fluctuations and electron-electron
correlations.\ 

Experimentally, there are some evidences for the existence of polarons\cite
{Bishop1}\cite{Billinge}{\it \ }\cite{deTeresa}\cite{Hennion}. But their
nature, their extension as a function of temperature and their dependence on
doping is not clear. In this paper, we will focus on the low doping regime,
close to the parent compound $LaMnO_{3}$, to look at the progressive
establishment of ferromagnetism and metallicity and to study the polaronic
effect.

For small doping concentration, several magnetic structures have been
observed at low temperature.\ In $La_{1-x}Sr_{x}MnO_{3}$\cite{Tokura (1994)}%
, a canted magnetic structure is present for $x\lesssim 0.08$, a
ferromagnetic insulator for $0.08\lesssim x\lesssim 0.16$ and a
ferromagnetic metal above. In $La_{1-x}Ca_{x}MnO_{3}$\cite{Schiffer}, the
compound has been reported as a ferromagnetic insulator for $x\lesssim 0.15$
and a ferromagnetic metal above. However, a different magnetic structure has
been observed by H.\ Hennion {\it et al}\cite{Hennion}: they found that, for 
$x=0.05$ and $x=0.08$, the compound is antiferromagnetic.\ For the parent
compound $LaMnO_{3}$\cite{Wollan}\cite{cristallo}, a variety of magnetic and
crystallographic structures have been observed due to off-stoichiometry on
the Lanthanum as well as the Manganese sites: antiferromagnetic, canted,
spin glass, ferromagnetic insulators or even ferromagnetic metals . These
examples are an illustration that the magnetic phase diagram depends on the
doping element as well as off-stoichiometry effects.

The samples studied here are doped in $Mn^{4+}$ through off stoichiometry in
Lanthanum. Their stoichiometry and crystallographic structure have been
determined by neutron diffraction. We have investigated the magnetic and
electric properties of two samples with different stoichiometries: 9\% and
21\% of $Mn^{4+}$. We will show how the magnetic phase diagram evolves as
doping is increased (from a canted magnetic structure to a ferromagnet), how
it influences their electric transport (both samples are insulators down to
low temperature) and their magnetoresistivity. We will present evidences for
polarons in both samples at high temperature (above the magnetic ordering
temperature) and we will discuss their role in the magnetic and electric
properties of the compound as a function of doping. Comparison with other
manganites with different doping element will be presented.

\section{Samples preparation and characterization}

Polycrystalline samples were prepared through a conventional solid state
reaction.\ The starting materials were $La_{2}O_{3}$(calcined 4N pure) and
dried $MnO_{2}$. Stoichiometric amounts of these powders were mixed and
heated at $900{{}^{\circ }}C$ for 12 hours in alumina crucibles. After
regrinding, the black powders were pellet pressed and sintered in an oxygen
gas flow during 24 hours at $950{{}^{\circ }}C$.\ Additional treatments at $%
950{{}^{\circ }}C$ under flowing oxygen or purified argon have been carried
leading to different off-stoichiometry. In each case, the samples were
cooled down at a rate of $100{{}^{\circ }}C$ per hour. An X ray diffraction
pattern was performed on each sample and analyzed using a Rietveld analysis.
The results for two typical samples LMO94R and LMO104RO3 are presented in
table I. LMO94R is orthorhombic whereas LMO104RO3 is rhombohedral. Presence
of $Mn_{3}O_{4}$ was detected in LMO94R.

Neutron diffraction data were collected on the high resolution
diffractometer D2B at ILL, Grenoble, to determine whether these samples had
a different cationic or oxygen stoichiometry.\ The structure has been
refined by the Rietveld method, using the Pnma space group for sample LMO94R
and R-3C for sample LMO104RO3. From the refined values of the occupancy
factors, it has be seen that these samples present La deficiency: $%
La_{0.97}MnO_{3}$ and $La_{0.93}MnO_{3}$ respectively (see table I). This
means that the exact formula is $La_{1-\delta }MnO_{3}$. The samples then
contain $Mn^{4+}$ ions with the concentration $3\delta $.

\section{Magnetic properties}

Off stoichiometry in $LaMnO_{3}$ has been widely reported \cite{cristallo},
and is known to modify the crystallographic and magnetic structure. Indeed, $%
La_{0.97}MnO_{3}$ and $La_{0.93}MnO_{3}$ have a very different magnetic
behavior. To characterize them, AC susceptibility and low field DC
magnetization measurements were used in the temperature range 4K-350K. The
field dependence of the magnetization was recorded with an extraction method
in a field up to 20 T at different temperatures, in the range 4K-300K. The
electrical resistivity was also measured in the same field and temperature
range using a DC technique for resistance up to $10^{7}$Ohm. The powdered
samples were then pellet pressed and silver paste was used for electric
contacts.

La$_{0.93}$MnO$_{3}$ is ferromagnetic below $T_{C}=170K$ (figure 1).\ Here,
the ordering temperature is taken at the inflection point on the
susceptibility curve. At 4K, the magnetization saturates in a field of 1
Tesla and remains constant up to 20 Tesla with a value $M_{s}=3.2\mu _{B}$.\
This value should be compared to the expected one.\ For both $Mn^{3+}$ and $%
Mn^{4+}$, the orbital momentum is quenched so that the magnetic moment
reduces to the spin contribution $gS\mu _{B}$ where S is the spin of the ion
(3/2 for $Mn^{4+}$ and 2 for $Mn^{3+}$) and g is the gyromagnetic factor
(g=2 for both). For 1-x $Mn^{3+}$ ions and x $Mn^{4+}$ ions, it becomes $%
(4-x)\mu _{B}$. Taking the value of x from the neutron diffraction results,
we get $3.79\mu _{B}$ , a value higher than observed.\ This discrepancy may
have two different origins. Either it arises from the presence of disorder
in the sample, although $M_{s}$ is not affected by a magnetic field up to 20
Tesla. No spin glass behavior has been detected (the ordering temperature
does not depend on the measurement frequency; the magnetization at low
temperature does not depend on time).\ More probably, this overestimation
reveals that the simple calculation for the saturation moment is affected by
collective effects and/or spin orbit coupling.

Above the ordering temperature, the system behaves quite differently from
what is expected: the magnetic susceptibility does not follow a Curie-Weiss
law. To have a better understanding of the paramagnetic regime, we have
analyzed the magnetization curves as a function of field within a mean field
approximation model.\ The internal field $H_{i}$ seen by the magnetic
moments has three contributions: the applied field $H_{a}$, the
demagnetization field $H_{d}$ and the molecular field $H_{m}$ due to the
magnetic coupling. In a mean field approximation, this molecular field can
be described by a single parameter $\lambda $ such that $H_{m}=\lambda M\,\,$%
where M\ is the magnetization. The demagnetization field is given by $%
\,H_{d}=nM$ where n is the demagnetization factor.\ Our powdered samples may
be considered as spheres for which $n=1/3$\thinspace .\ This demagnetization
field is quite small compared to the applied field ($\mu _{0}H_{d}\simeq
0.3T $ when the magnetization is at its maximum value $M=3.2\mu _{B}$
whereas $\mu _{0}H_{a}$varies up to 20T).\ We have plotted, in the insert of
figure 2, the magnetization as a function of $H_{i}/T$. All the
magnetization curves collapse on a single curve in the temperature range
100K-300K if the mean field parameter is taken as temperature dependent (see
figure 2).\ This is in agreement with the susceptibility data which do not
follow the mean field Curie-Weiss law. The curve $M\left( H_{i}/T\right) $
can be fitted with a single Brillouin function from which we deduce an
effective magnetic moment $\mu _{eff}$. We get $\mu _{eff}=5.8\mu _{B}$. In
a mean field approximation, for one type of magnetic ions, $\mu _{eff}$ is
given by :$\mu _{eff}^{2}=xg^{2}S(S+1)\mu _{B}^{2}$ where x is the fraction
of magnetic ions per formula unit, g is their gyromagnetic factor and S
their spin.\ $La_{0.93}MnO_{3}$ contains $Mn^{3+}$and $Mn^{4+}$ ions.\ We
can treat them as one type of ions with an average spin $S=2(1-x)+\frac{3}{2}%
x$ with x = 0.21, which lead to $\mu _{eff}=4.68\mu _{B}$\smallskip $.\;$Or
we can treat them as separated magnetic systems with the same ordering
temperature. then $\mu _{eff}^{2}=\mu _{eff1}^{2}+\mu _{eff2}^{2}$ and $\mu
_{eff}=4.44\mu _{B}$. In both cases, the observed value is much bigger.\ It
is the signature of clusters of $Mn^{4+}$ and $Mn^{3+}$.

$La_{0.97}MnO_{3}$ behaves quite differently. A peak in the AC
susceptibility is observed at 118K (figure 3, upper part). The additional
peak at 43K is related to $Mn_{3}O_{4}$. The DC susceptibility shows
hysteresis below 118K revealing the presence of ferromagnetic domains: it
has a more ferromagnetic like behavior (figure 3, lower part). The
magnetization curves as a function of field at low temperature present a
spontaneous magnetization followed by a high field susceptibility (figure
4). These features are characteristics of a canted magnetic structure. This
was confirmed by neutron diffraction data. At $2.5K$, the ferromagnetic
moment is $1.48\mu _{B}$ and the antiferromagnetic moment $2.93\mu _{B}$, so
that the canting angle is $130{{}^{\circ }}$. The measured high field
susceptibility, $\chi \simeq 8.75\,10^{-2}\mu _{B}/T\simeq 1.74\times
10^{-2}S.I.$, is temperature independent from $4K$ up to $100K$ at least. We
have also observed some hysteresis up to $20T$. This hysteresis exists in a
limited temperature range: $T<80K$ (figure 5).

Above the ordering temperature, we have analyzed the magnetization curves in
terms of a mean field theory, similarly to $La_{0.93}MnO_{3}$. The molecular
field parameter is constant from $300K$\ until $150K$ (figure 2), which
suggests that a mean field approach is correct. The effective moment deduced
from a Brillouin fit is $\mu _{eff}=6.0\mu _{B}$\smallskip .\ For this
concentration also, clusters are present.\ The calculated mean field
ordering temperature is $T_{P}=114K$, a value very close to the observed
transition temperature $(T_{C}=118K)$.

\section{\protect\smallskip Electric transport}

Both samples remain insulating (figure 5) with activation energies of the
order of 0.16 eV. No anomalies or change of slope have been detected at the
ordering temperature.\ Application of a magnetic field reduces the magnitude
of the electric resistance.\ The resistance as a function of field has a
smooth variation for both samples. When the electric resistivity, $\rho $,
is plotted as a function of the magnetization $M$, quite different behavior
are observed for each sample. The canted sample, $La_{0.97}MnO_{3}$,
presents a universal law for $\rho \left( M\right) $ down to 120 K : $\rho
\left( M\right) /\rho \left( 0\right) =1.25\left( M/M_{s}\right) ^{2}$(see
figure 6).\ Measurements were restricted to $T\geq 120K$ because of the high
resistivity. \ At 120K, around the ordering temperature, an additional
process contributes to the electric transport at low magnetization.\ It is
probably related to magnetic domains.

The ferromagnetic sample behaves very differently. The square dependence of
the magnetoresistance on the magnetization is no longer observed. Actually,
no universal dependence can be deduced. Granular effects are probably too
important.

\section{Discussion}

The fundamental electronic structure of these mixed valence oxides have been
reviewed by J.B. Goodenough \cite{Goodenough}. $LaMnO_{3}$ involves $Mn^{+3}$
ions in the $t_{2g}^{3}e_{g}^{1\text{ }}(S=2)$ configuration. These are
Jahn-Teller ions. $LaMnO_{3}$ is well described by a localized electron
model. In the Mn-O-Mn plane, the half filled $e_{g}$ orbital couples to an
empty $e_{g}$ orbital, leading to a ferromagnetic coupling. In the
perpendicular direction, coupling between half filled $t_{2g}$ orbitals
leads to antiferromagnetism. $LaMnO_{3}$ has therefore a layered
antiferromagnetic structure.

When doping by substituting the trivalent La ion by divalent ions such as
Ba, Ca, Sr, ... (or adding oxygen), $Mn^{4+}$ ions are introduced. They are
in the $t_{2g}^{3}e_{g}^{0}$ $(S=3/2)$ configuration. Above a critical
concentration of the order of 0.1, the $e_{g}$ electrons are delocalized.
Due to their intrinsic spin and strong correlation with the localized $%
t_{2g}^{3}$ electrons, they hop from a $Mn^{+3}$ site to a $Mn^{+4}$ site
having the same spin orientation. This so-called double exchange gives rise
to a ferromagnetic metallic state. The transfer energy between two ionic
sites having spins making an angle $\theta _{ij}$ is expressed by [12]: $%
t_{ij}~=~b_{ij}~cos(\theta _{ij}/2)$. This introduces a canting of the spin; 
$\theta $ increases with $Mn^{4+}$ content until complete ferromagnetism is
reached [9], around $x\simeq ~0.3$~.The canted state is characterized by a
spontaneous magnetization $M_{s}$ which is related to the canting angle $%
\theta $:

$M_{s}=Icos(\theta /2)$ (1),

$cos(\theta /2)=bx/4\left| J\right| S^{2}$ (2)

and by a high field susceptibility which is roughly equal to the
perpendicular susceptibility of the antiferromagnetic parent compound $%
LaMnO_{3}$:

$\chi _{_{\bot }}=I^{2}/4\left| J\right| \gamma _{0}S^{2}N$ (3).

Here $I/2$ is the sublattice magnetization, $x$ the $Mn^{4+}$ concentration, 
$J$ the antiferromagnetic coupling, $b$ the double exchange coupling, $N$
the number of magnetic ions per unit cell $\left( N=1\right) $, $S$ their
spin and $\gamma _{0}$ the number of their neighbors in the adjacent layers $%
(\gamma _{0}=2)$.

The above features have been derived by P.G. de Gennes\cite{de Gennes}
assuming a pure canted spin arrangement, such that the spins form two
sublattices separated by an angle $\theta $. However, J. Inoue {\it et al} 
\cite{Inoue} have pointed out that the pure canted state should be more
stable close to $x=1$ and a spiral state close to $x~=~0$.

In both cases, the spin in one set of Mn-O plane is rotated by an angle from
the adjacent Mn-O plane. The distinction between these two arrangements is
the following: in the pure canted state, there are two sublattices making an
angle $\theta $ with each other. The magnetic periodicity is therefore twice
the lattice periodicity in the direction perpendicular to the Mn-O planes.
In the spiral state, the successive planes make an angle $\theta $ , $%
2\theta $, $3\theta $, ... with respect to a reference plane. The
periodicity can be large, $\theta /2\pi $ times the lattice periodicity and
incommensurability effects may occur. The model by P.G. de Gennes should not
be substantially affected whether the spin arrangement is of the pure
canting or spiral type.

E.O.\ Wollan {\it et al} [3] have reported evidences for pure canted spins
at low concentration. Their data agree with equation (2). Our experimental
results for $La_{0.97}MnO_{3}$ agree quite well also with such a picture:
the magnetization curves present a spontaneous magnetization followed by a
temperature independent susceptibility. The spontaneous magnetization is in
agreement with the value of the canting angle deduced from neutron
diffraction data. From the high field susceptibility, $\chi \approx
1.74\times 10^{-2}S.I.$ , we can estimate the antiferromagnetic coupling,
using the average value for S $\left( S=1.96\right) $: $J\approx
-4.0K\approx -0.36meV$. This value should be compared to the one deduced
from spin wave measurements: in pure $LaMnO_{3}$, $J\approx -0.58meV$ \cite
{Hirota1}\cite{Moussa}; in $La_{0.95}Ca_{0.05}MnO_{3}$ , it is reduced to $%
J\approx -0.38meV\ $\cite{Hennion}, and in $La_{0.95}Ca_{0.05}MnO_{3}$, to $%
J\approx -0.28meV$;$\;$in $La_{0.95}Sr_{0.05}MnO_{3}$ , it is $J\approx
-0.20meV\ $\cite{Hirota2}.\ Our results are in good agreement with those of $%
La_{1-x}Ca_{x}MnO_{3}$.\ The antiferromagnetic interaction is reduced in the
similar way when doping is introduced through Lanthanum vacancies or through
Calcium doping. This is not very surprising considering the great mismatch
between the size of Lanthanum ions $\left( r=0.122nm\right) $ and the one of
Calcium ions $\left( r=0.106nm\right) $. On the contrary, doping through
Strontium ions is different since its ionic radius is bigger $\left(
r=0.126nm\right) $. The overlap of the oxygen and manganese orbitals is
modified in a different way.

Using equation (2), we can extract the double exchange coupling: $b\approx
290K$.\ In term of energy, the double exchange coupling contributes to an
amount of $E_{d1}=-4bx\approx -104K$ and $E_{d2}=-2\frac{b^{2}x^{2}}{4\left|
J\right| S^{2}}\approx -22K$; the first term refers to the double exchange
energy within one Manganese plane and the second term to the double exchange
energy from one plane to the adjacent planes. The antiferromagnetic
contribution from one plane to the adjacent ones is $E_{AF}=2\left| J\right|
S^{2}\cos \theta _{0}\approx -20K$.\ The ferromagnetic contribution arising
from the indirect exchange coupling between manganese ions in the same plane
is given by $E_{F}=-4J^{\prime }S^{2}$ where J' is the ferromagnetic
indirect coupling. In $La_{0.95}Ca_{0.05}MnO_{3}$, $J^{\prime }\approx 12K$ 
\cite{Hennion}, which leads to $E_{F}\approx -184K$. In term of energy, the
double exchange contribution has the same order of magnitude as the
antiferromagnetic contribution $\left( -22K\text{ and }-20K\text{
respectively}\right) $.\ It is not strong enough to establish complete
ferromagnetism. The slope of the function $\cos \theta _{0}$ as a function
of $x$ is similar to what is observed in $La_{1-x}Ca_{x}MnO_{3}$ \cite{de
Gennes}. This confirms the great similitude between $La_{1-\delta }MnO_{3}$
and this system, in term of exchange interactions at least.

At low temperature, a high field hysteresis is present below 80K (see figure
4). Repeated magnetization curves show a random behavior between two extrema
curves. We propose that this hysteresis finds its origin in the different
canted structures that are possible with a fixed canting angle i.e. spiral
or pure canted or mixture.\ It is not a field induced phase transition but
it reflects rather the presence of different metastable configurations for
the spins. However we did not detect any time dependent effects on the
magnetization, at least on a time scale of a couple of hours. The
disappearance of this hysteresis above 80 K can be explained by thermal
activation effects.\ It can also be explained by the occurrence of a charge
ordered phase present only at low temperature.

P.G.\ de Gennes \cite{de Gennes} has calculated the characteristic ordering
temperatures $T_{C}$ and $T_{P}$ where $T_{C}$ is the ordering temperature
and $T_{P}$ the paramagnetic Curie point defined through the asymptotic form
of the susceptibility $\chi =C/\left( T-T_{C}\right) $. He found that they
should coincide: $k_{B}T_{C}=k_{B}T_{P}=\frac{2}{3}\left[ \left( 2\left|
J\right| +4J^{\prime }\right) S^{2}-\frac{4}{5}bx\right] $, where J' takes
into account the modification arising from the Zener carrier (J' increases
with x as has been observed in \cite{Hennion}). Taking the values determined
above for $J$, $J\prime $ and $b$, we get: $T_{C}=T_{P}\approx 130K$.
Experimentally, we do observe $T_{C}\simeq T_{P}$ with a slightly smaller
value $(114K)$.\ The treatment by P.G.\ de Gennes seems quite appropriate.

At higher temperature, the paramagnetic regime is however not conventional:
it is described by an effective magnetic moment which is much bigger than
expected. This is the signature of magnetic clusters. Many combinations of $%
Mn^{3+},Mn^{4+}$ can give rise to $\mu _{eff}=6.0\mu _{B}$ so that it is
difficult to extract the exact size of these clusters. The magnetic field
does not affect their size nor their number, since we observe a single
Brillouin function as a function of $H_{i}$.\ Temperature effects are only
visible close to $T_{C}$: the mean field parameter starts decreasing for $%
T\leq 1.2T_{C}$.\ This reflects the onset of fluctuations.

What is the origin of these clusters? In these materials, presence of
polarons has been detected\cite{Bishop1}\cite{Billinge}{\it \ }\cite
{deTeresa}\cite{Hennion}.\ It depends on the doping concentration as well as
the nature of the doping element. For instance, deviation of the
susceptibility from a Curie law is usually observed above $T_{C}$. In $%
La_{0.66}Ca_{0.33}MnO_{3}$ \cite{deTeresa}, it occurs below $1.8T_{C}$ and
is accompanied by the formation of magneto-elastic polarons. Their coherence
length increases when the temperature is lowered and eventually diverges at $%
T_{C}$. It also increases with magnetic field with a typical value at zero
field of $1.2nm$.$\;$Above $1.8T_{C}$, a Curie-Weiss law is recovered with
an effective moment corresponding to isolated manganese ions. $%
La_{0.97}MnO_{3}$ is an insulator in the whole temperature range.\ The
polaronic effect is different: magnetic clusters are present at much higher
temperature (until $2.5T_{C}$ at least) with a size remaining constant in
the whole temperature and field range studied. Deviation from a Curie law
occurs only when $T<1.2T_{C}$. the magnetic polaron is more strongly
established in $La_{0.97}MnO_{3}$ than in $La_{0.66}Ca_{0.33}MnO_{3}$. The
magneto-elastic interaction which is responsible for its formation is much
stronger. This polaronic effect depends on the doping concentration.\ In $%
La_{1-x}Ca_{x}MnO_{3}$, S.J.Billinge {\it et al} \cite{Billinge}{\it \ }
have observed lattice polarons above the ferromagnetic transition for $%
x=0.25 $ and $x=0.21$, but none for $x=0.12$ which does not present an
insulating to metallic transition like the others at the ferromagnetic
transition. This is in contrast to our results where we observe polarons in $%
La_{0.97}MnO_{3}$ which is even less doped; but this is consistent with the
idea that the polarons are 'stronger' in $La_{1-\delta }MnO_{3}$ than in $%
La_{1-x}Ca_{x}MnO_{3}$.\ 

Several theoretical treatments of the polaronic effects have been proposed.
A.J.Millis{\it \ et al} \cite{Millis (1996)} conclude that, for intermediate
doping, strong electron-phonon coupling localizes the conduction electrons
as polarons for $T>T_{C}$ and that the effect is turned off at $T_{C}$.
H.R\"{o}der{\it \ et al }\cite{Roder1996}{\it \ }arrive to the same
conclusion . They were able to reproduce the doping dependence of the
ferromagnetic transition $T_{C}$ and found that, for $T>T_{C}$, in the
dilute limit, small magnetopolarons are formed and comprise a localized
charge surrounded by a spin cloud. They have shown that the size of the
polaron grows as the ordering temperature is approached from above, in
agreement with the experimental results on $La_{0.66}Ca_{0.33}MnO_{3}$ \cite
{deTeresa}. This lattice polaron picture remains controversial: C.M. Varma 
\cite{Varma} proposed rather the idea of spin polarons due to random hoping
driven by thermal fluctuations and electron-electron correlation.\ He found
also that the magnetic susceptibility should be enhanced above $T_{C}$ with
a temperature dependent effective moment, in a first approximation at least.

All these theoretical works are only concern with the ferromagnetic metallic
phase and do not seem appropriate for $La_{0.97}MnO_{3}$ . Indeed, in this
sample, the size of the polarons is not temperature dependent in the
paramagnetic regime, which is major difference compare to the ferromagnetic
metallic phases. But we may have observe precursor effects.

Does this clustering persist in the ordered phase? At low temperature, the
magnetization measurements agree quite well with de Gennes's theory of
canted spins. However this cannot rule out the existence of clusters; as was
pointed out by P.G.\ de Gennes \cite{de Gennes}, the same magnetic behavior
is expected both for long range order or bound states. In\ a sample with the
same doping level, M. Hennion {\it et al} \cite{Hennion} have observed, at
low temperature, magnetic excitations that they have interpreted as magnetic
polarons. They disappear above the ordering temperature.\ It is quite
different in our sample where the magnetic polarons exist in the
paramagnetic phase.

This different behavior should be accounted for because of the different
nature of doping in both samples: for $La_{0.97}MnO_{3}$, vacancies on the
lanthanum is the doping factor whereas in $La_{1-x}Ca_{x}MnO_{3}$, doping is
introduced by replacing lanthanum ions by the smaller calcium ions. The
lattice mismatch is greater in $La_{1-\delta }MnO_{3}$ than in $%
La_{1-x}Ca_{x}MnO_{3}$.\ The magnetic order is affected: $La_{0.97}MnO_{3}$
has a canted structure with $\theta \simeq 130{{}^{\circ }}$, while $%
La_{0.92}Ca_{0.08}MnO_{3}$ is antiferromagnetic $(\theta \simeq 170{%
{}^{\circ }})$\cite{Hennion}.The polaronic effect is also affected. $%
La_{0.95}Ca_{0.05}MnO_{3}$ and $La_{0.95}Sr_{0.05}MnO_{3}$, have the same
doping level (5\% of $Mn^{4+}$) and the same antiferromagnetic magnetic
structure. Magnetic excitations have been detected at low temperature for
the former \cite{Hennion} and none for the later \cite{Hirota2}.These
results confirm that the strength of the polarons depends crucially on the
nature of the doping element due to the modification of the electron-phonon
coupling as well as the magnetic interactions.

If polarons are present, the resistivity is expected to follow the variable
range hoping law.\ We only observe an activated behavior as a function of
temperature. This is consistent with the fact that the polarons remains
constant in size.\ We observe a simple law for $\rho \left( M\right) $: $%
\rho \left( M\right) /\rho \left( 0\right) =C\left( M/M_{s}\right) ^{2}$
with $C\simeq 1.25$ (see figure 6). Such a law has been predicted by
N.Furukawa \cite{Furukawa}. \ His calculation is based on spin
fluctuations.\ He shows that the constant $C$ depends both on the doping
level and the Hund coupling between the $e_{g}$ electrons and the localized $%
S=3/2$ spins. Numerical values for $C$ are given for higher doping
concentration than $La_{0.97}MnO_{3}$ but we can extrapolate to lower
concentration: the tendency is to get a higher value of $C$ when the doping
is decreased, at least in the strong Kondo coupling limit, as it is the case
in these compounds \cite{Furukawa2}. In $La_{0.97}MnO_{3}$, we get a value
of $C$ plausible only in the weak coupling limit.\ This is in contrast to
what is observed in $La_{1-x}Sr_{x}MnO_{3}$ \cite{Tokura (1994)} where the
vale of $C$ is about $4$ for $0.15\leq x\leq 0.2$, and decreases to $2$ for $%
x=0.3$ and about $1$ for $x=0.4$.

So far, we have discussed the properties of $La_{0.97}MnO_{3}$. How are they
affected when more vacancies are introduced? First of all, the system
becomes ferromagnetic (with complete saturation at $4K$), as expected from
double exchange theory.\ It is remarkable to notice, though, that $%
La_{0.93}MnO_{3}$ never achieves a metallic conductivity. Indeed no anomaly
in the resistivity has been observed down to 50K even in high magnetic
field. A similar ferromagnetic insulating phase has been observed in $%
La_{0.90}Sr_{0.10}MnO_{3}$ \cite{Urushibara} and in $LaMnO_{3.13}$ \cite
{Goodenough}.\ At first sight, this is in contradiction with the idea of
double exchange where ferromagnetism is induced by electron hoping and
therefore requires an insulator to metal transition when long range
ferromagnetic order is established.\ The existence of a ferromagnetic
insulating phase, as in $La_{0.93}MnO_{3}$, is a clear evidence that not all
the Zener electrons responsible for ferromagnetism take part in the
conduction process. Electron localization occurs. This is also observe at
higher $Mn^{4+}$ content:\ in $La_{0.7-x}Y_{x}Ca_{0.3}MnO_{3}$ thin films 
\cite{Fontcuberta}, with a constant doping level (30\% of $Mn^{4+}$), the
insulator to metal transition is shifted to lower temperature as a function
of yttrium doping and, at the same time , the level of resistivity in the
ferromagnetic phase is increased by several order of magnitude.

\smallskip This localization has been explained by R. Allub {\it et al}\cite
{Alascio}. They have looked at the effect of localized states introduced by
disorder. Their model was successful to account for the change in
resistivity in $La_{1-x}Sr_{x}MnO_{3}$ \cite{Tokura (1994)} where the system
is ferromagnetic and insulating for x 
\mbox{$<$}%
0.125. In $La_{1-\delta }MnO_{3}$, the insulating state persists to much
higher concentration (for $La_{0.93}MnO_{3}$, $x=0.21$). Another origin for
this insulating behavior may be found in the presence of polarons persistent
in the ferromagnetic phase. It has been shown that, in the metallic
ferromagnet \cite{Billinge}\cite{deTeresa}, the polarons disappears below $%
T_{C}$. In $La_{0.93}MnO_{3}$, similarly to $La_{0.97}MnO_{3}$, the
polaronic effect is much stronger and may remain effective at low
temperature.\ It will localize the electrons. However, the ferromagnetic
correlation length will have to be greater than the polaron size to get a
fully ferromagnetic phase. More theoretical developments are required.\ 

Experimentally, in the paramagnetic regime, we observe a smaller effective
moment in $La_{0.93}MnO_{3}$ than in $La_{0.97}MnO_{3}$ (from $6.0$ to $%
5.8\mu _{B}$).\ This suggests that the polaron size has decreased. Similar
results have been observed by J.T\"{o}pfer {\it et al} \cite{cristallo} in $%
La_{1-\delta }Mn_{1-\delta }O_{3}$. Indeed, we would expect, for even higher
vacancy content, to reduce further the size of the polarons to rich a single
ion behavior with, at the same time, an enlargement of the critical regime
and occurrence of an insulator to metal transition with huge
magnetoresistance. However, in this $La_{1-\delta }MnO_{3}$ series, the
occurrence of a ferromagnetic metallic like phase is reduced as a function
of the $Mn^{4+}$ content compare to $La_{1-x}Sr_{x}MnO_{3}$ or $%
La_{1-x}Ca_{x}MnO_{3}$. The metallic phase may even not exist, as it is the
case for $\Pr_{1-x}Ca_{x}MnO_{3}$ \cite{NdCaMnO}. Here, an additional
process is involved: charge ordering occurs.

$\smallskip $

\section{Conclusions}

We have studied the magnetic and electric properties of $La_{1-\delta
}MnO_{3}$ for two values of $\delta $: $\delta =0.03$ and $\delta =0.07$
corresponding to 9\% and 21\% of $Mn^{4+}$respectively.\ Their properties
change dramatically. $La_{0.97}MnO_{3}$ has a canted magnetic structure at
low temperature $(\theta \simeq 130{{}^{\circ }}$ and $T_{C}=118K)$. Its
magnetic behavior fits quite well with de Gennes theory of canted spins.\ We
have extracted the values of the double exchange interaction $\left(
b\approx 290K\right) $ and the antiferromagnetic indirect exchange $\left(
J\approx -4.0K\right) $. In the paramagnetic regime, magneto-elastic
polarons are present with an effective moment of $\mu _{eff}=6.0\mu _{B}$.\
They size is not affected by temperature for $1.2T_{C}<T<2.5T_{C}$, nor by
magnetic field $(B<20T)$ contrary to what is observed in the metallic
ferromagnet $La_{0.66}Ca_{0.33}MnO_{3}$.\ For higher vacancy content, $%
\delta =0.07$, polarons are still present but smaller in size; their
effective moment is $\mu _{eff}=5.8\mu _{B}$.\ The system becomes fully
ferromagnetic thanks to double exchange, but the compound remains
insulating, even in the ferromagnetic phase.

The Grenoble High Magnetic Field Laboratory is ''Laboratoire
conventionn\'{e} \`{a} l'Universit\'{e} Joseph Fourier de Grenoble''.

\section{Figures}

{\bf Figure 1}: Temperature dependence of the magnetic susceptibility in $%
La_{0.93}MnO_{3}$.

upper part: AC susceptibility

lower part: DC susceptibility in a field of 5mT (zero field cooling and
field cooling curves). Insert: Magnetization as a function of field at 4K.

\smallskip

{\bf Figure 2}: Mean field parameter $\lambda $ as a function of temperature
for $La_{0.97}MnO_{3}$ and $La_{0.93}MnO_{3}$.\ Insert: Magnetization as a
function of $g\mu _{B}H_{i}/kT$ for $La_{0.93}MnO_{3}$.

\smallskip

{\bf Figure 3}: Temperature dependence of the magnetic susceptibility in $%
La_{0.97}MnO_{3}$.

upper part: AC susceptibility

lower part: DC susceptibility in a field of 5mT (zero field cooling and
field cooling curves).

\smallskip

{\bf Figure 4}: Field dependence of the magnetization in $La_{0.97}MnO_{3}$%
.\ Insert: Hysteresis field $H_{h}$ as a function of temperature.

\smallskip

{\bf Figure 5}: Temperature dependence of the electrical resistivity in $%
La_{0.93}MnO_{3}$ and $La_{0.97}MnO_{3}$ without and with a 20 Tesla
magnetic field.

\smallskip

\smallskip {\bf Figure 6}: Resistivity as a function of magnetization for $%
La_{0.97}MnO_{3}$ .\ \newpage \vspace{0pt}

\begin{tabular}{|c|c|c|c|c|}
\hline
sample & formula & $Mn^{4+}$ content & space group & cell parameters \\ 
\hline
LMO94R & $La_{0.97}MnO_{3}$ & 9\% & Pnma & 
\begin{tabular}{c}
$a=0.56290nm\,\,\,$ \\ 
$b=0.77248nm$ \\ 
$c=0.55410nm$%
\end{tabular}
\\ \hline
LMO104 & $La_{0.93}MnO_{3}$ & 21\% & R-3C & 
\begin{tabular}{c}
$a=0.55245nm$ \\ 
$c=1.33433nm$%
\end{tabular}
$\,$ \\ \hline
\end{tabular}


\begin{references}
\bibitem{Jin}  S.T.Jin, T.H.Tielfel, M.McCormack,R.A.Fastnacht, R.Ramesh,
L.H.Chen, Science {\bf 264}, 413 (1994)

\bibitem{Asamitsu (nature)}  A.Asamitsu, Y.Moritomo, Y.Tomioka, T.Arima,
Y.Tokura, Nature {\bf 373}, 407 (1995)

\bibitem{Tomioka (PRL)}  Y.Tomioka, A.Asamitsu, Y.Moritomo, H.Kuwahara,
Y.Tokura, Phys.Rev.Lett. {\bf 74}, 5108 (1995)

\bibitem{Hwang (95)}  H.Y.Hwang, S-W.Cheong, P.G.Radaelli, M.Marezio,
B.Batlogg, Phys.Rev.Lett. {\bf 75}, 914 (1995)

\bibitem{Wollan}  E.O.Wollan, W.C.Koehler, Phys.Rev. {\bf 100}, 545 (1955)

\bibitem{Jonker}  G.H.Jonker, J.H.van Santen, Physica {\bf 16}, 337 (1950),
Physica {\bf 22}, 707 (1956)

\bibitem{Zener (1951)}  C.Zener, Phys.Rev. {\bf 82}, 103 (1951)

\bibitem{Anderson (1955)}  P.W.Anderson, H.Hasegawa, Phys. Rev. {\bf 100},
675 (1955)

\bibitem{de Gennes}  P.G.de Gennes, Phys. Rev. {\bf 118}, 141 (1960)

\bibitem{Kubo (1972)}  K.Kubo, N.Ohata, J.Phys.Soc.Japan {\bf 33}, 21 (1972)

\bibitem{Millis (1995)}  A.J.Millis, P.B.Littlewood, B.I.Shraiman,
Phys.Rev.Lett.\ {\bf 74}, 5144 (1995)

\bibitem{Bishop1}  for a review, see for instance A.R.\ Bishop, H.
R\"{o}der, cond-mat/9703148

\bibitem{Millis (1996)}  A.J.Millis, B.I.Shraiman, R. Mueller
Phys.Rev.Lett.\ {\bf 77}, 175 (1996)

\bibitem{Roder1996}  H.R\"{o}der, J.Zang, A.R.Bishop, Phys.Rev.Lett. {\bf 76}%
, 1356 (1996)

\bibitem{Varma}  C.M.Varma, Phys.Rev.B {\bf 54}, 7328 (1996)

\bibitem{Billinge}  S.J.L.Billinge, R.G.DiFrancesco, G.H.Kwei, J.J.Neumeier,
J.D.Thomson, Phys.Rev.Lett. {\bf 77}, 715 (1996)

\bibitem{deTeresa}  J.M.de Teresa, M.R.Ibarra, P.A.\ Algarabel, C.Ritter,
C.Marquina, J.Blasco, J.Garcia, A.del Moral, Z.\ Arnold, Nature {\bf 386},
256 (1997)

\bibitem{Hennion}  M.Hennion, F.Moussa, J.Rodriguez-Carvajal, L.Pinsard,
A.Revcolevschi, Phys.Rev.B, 56, {\bf 1} (1997)

\bibitem{Tokura (1994)}  Y. Tokura, A. Urushibara, Y. Moritomo, T. Arima, A.
Asamitsu, G. Kido, N. Furukawa, J.Phys.Soc. Jpn. {\bf 63}, 3931 (1994)

\bibitem{Schiffer}  P.Schiffer, A.P.Ramirez, W.Bao, S.-W.Cheong,
Phys.Rev.Lett. {\bf 75}, 3336 (1995)

\bibitem{cristallo}  see for instance J.T\"{o}pfer, J.B.Goodenough, J.Solid
State Chem. {\bf 130}, 117 (1997) and Q.Huang, A.Santoro, J.W.Lynn,
R.W.Erwin, J.A. Borchers, J.L.Peng, R.L.Greene, Phys.Rev.B {\bf 55}, 14987
(1997)

\bibitem{Goodenough}  J.B.Goodenough, {\it Progress in Solid State Chemistry}%
, edited by H. Reiss (Pergamon, New York 1971) vol5, chap.4, p325

\bibitem{Inoue}  J.Inoue, S.Maekawa, Phys.Rev.Lett. {\bf 74}, 3407 (1995)

\bibitem{Hirota1}  K.Hirota, N.Kane A.Nishizawa, Y.Endohko, J.Phys.Soc.Jpn. 
{\bf 66}, 3736 (1996)

\bibitem{Moussa}  F.Moussa, M.Hennion, J.Rodriguez-Carvajal, H.Moudden,
L.Pinsard, A.Revcolevski, Phys.Rev.B {\bf 54}, 15149 (1996)

\bibitem{Hirota2}  K.Hirota, N.Kaneko, A.Nishizawa, Y.Endoh, M.C.Martin,
G.Shirane, PhysicaB 237-238, {\bf 36} (1997)

\bibitem{Furukawa}  N. Furukawa, J.Phys.Soc.Jpn. {\bf 63}, 3214 (1994)

\bibitem{Furukawa2}  Y.Tokura, A.Urushibara, Y.Moritomo, T.Arima,
A.Asamitsu, G.Kido, N.Furukawa, J.Phys.Soc.Jpn. {\bf 63}, 3931 (1994)

\bibitem{Urushibara}  A.Urushibara, Y.Moritomo, T.Arima, A.Asamitsu, G.Kido,
Y.Tokura, Phys.Rev.B {\bf 51}, 14103 (1995)

\bibitem{Fontcuberta}  J.Fontcuberta, B.Martinez, A.Seffar, S.Pi\~{n}ol,
J.L.Garcia-Mu\~{n}oz, X.Obradors, Phys.Rev.Lett. {\bf 76}, 1122 (1996)

\bibitem{Alascio}  R.Allub, B.Alascio, Solid State Com. {\bf 99}, 613 (1996)

\bibitem{NdCaMnO}  Y.Tomioka, A.Asamitsu, H.Kuwahara, Y.Moritomo, Y.Tokura,
Phys.Rev.B {\bf 53}, R1689 (1996)
\end{references}
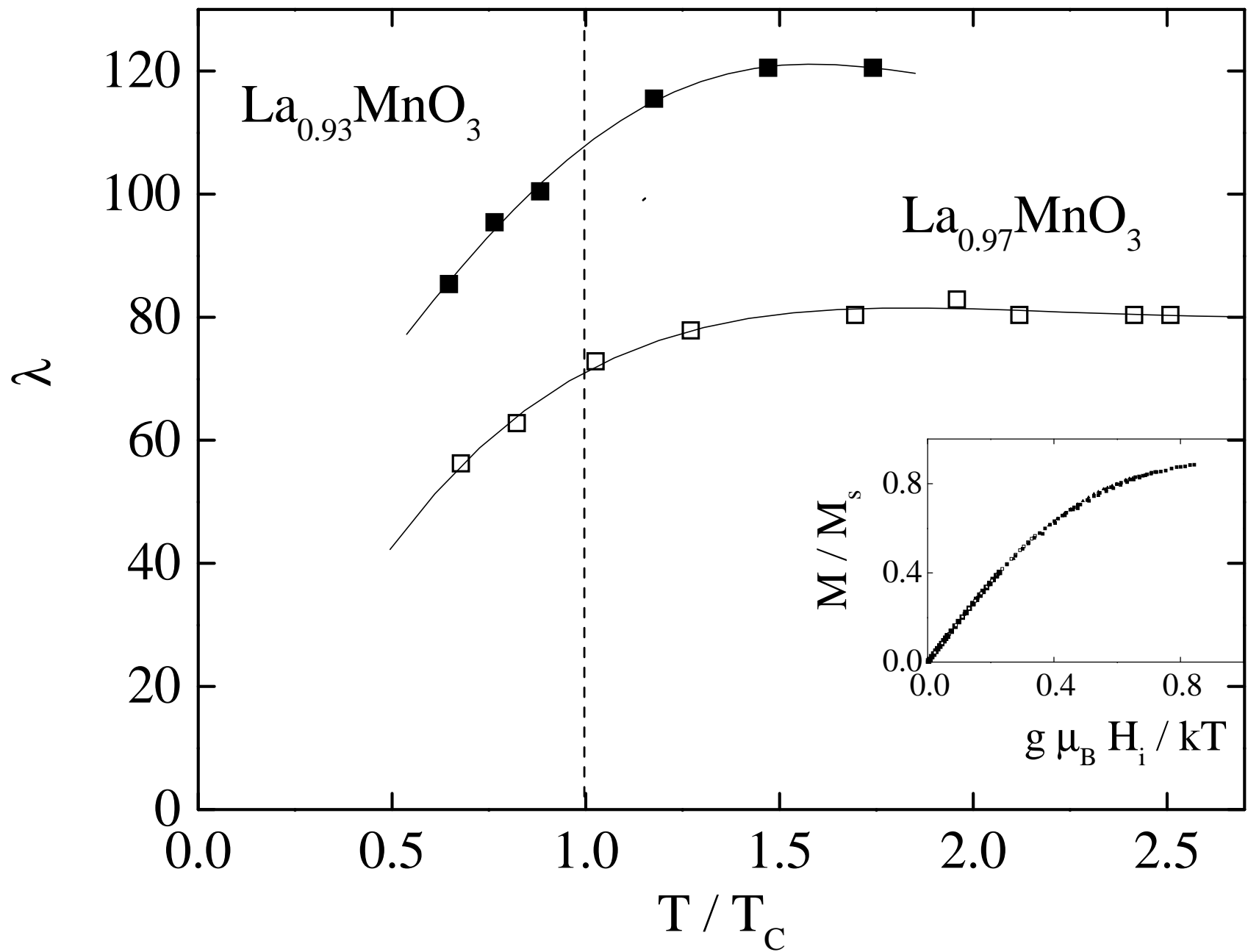
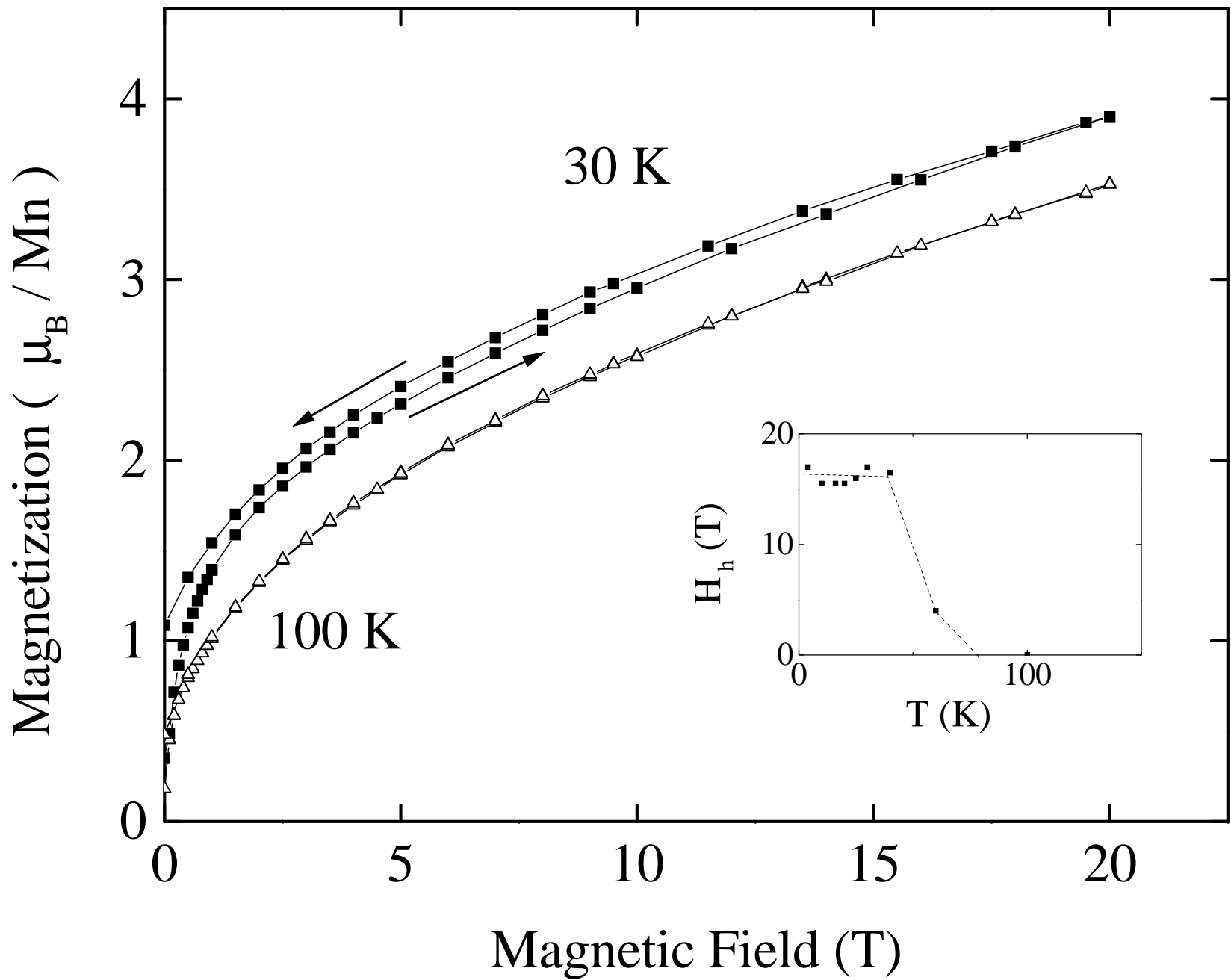
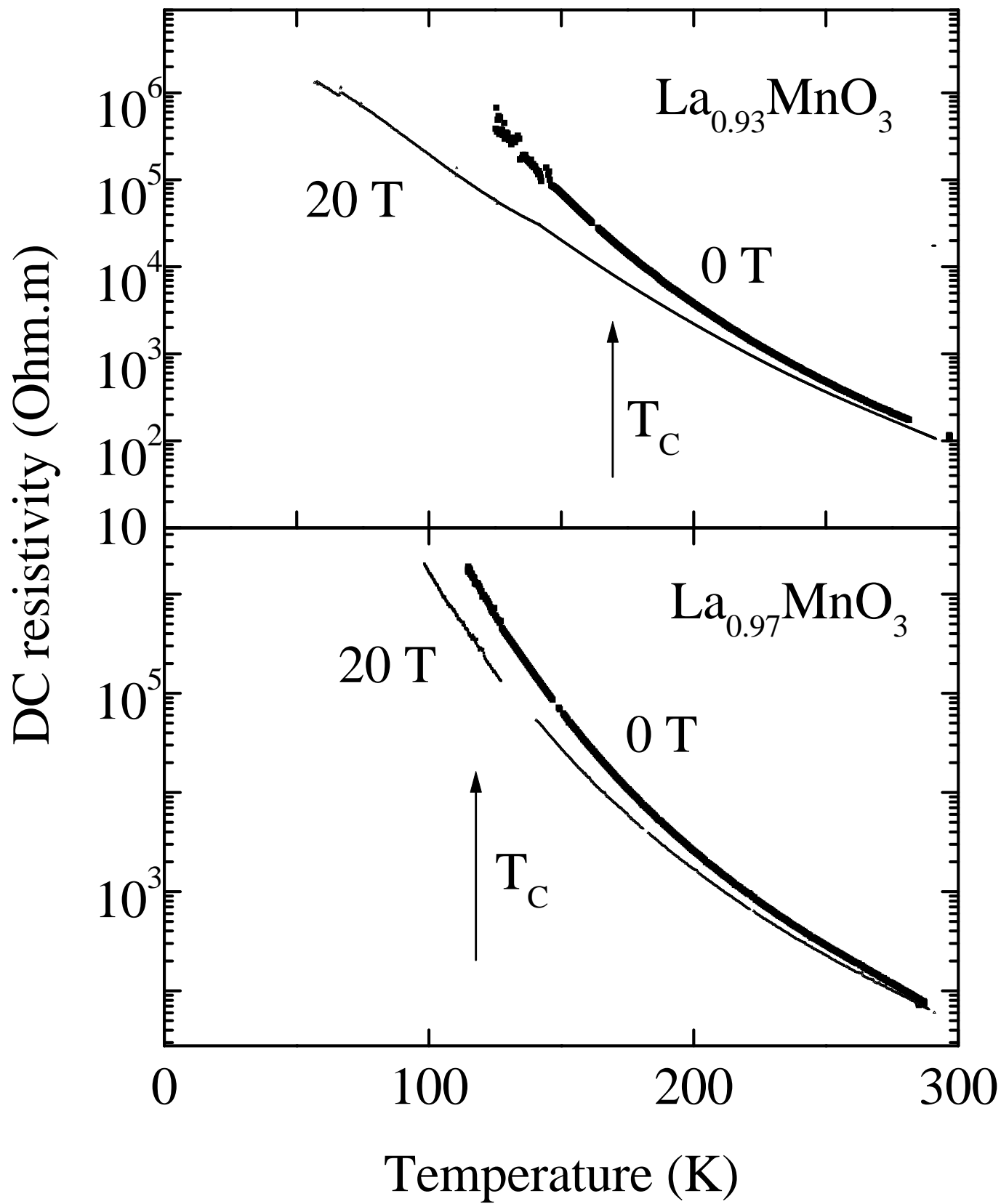
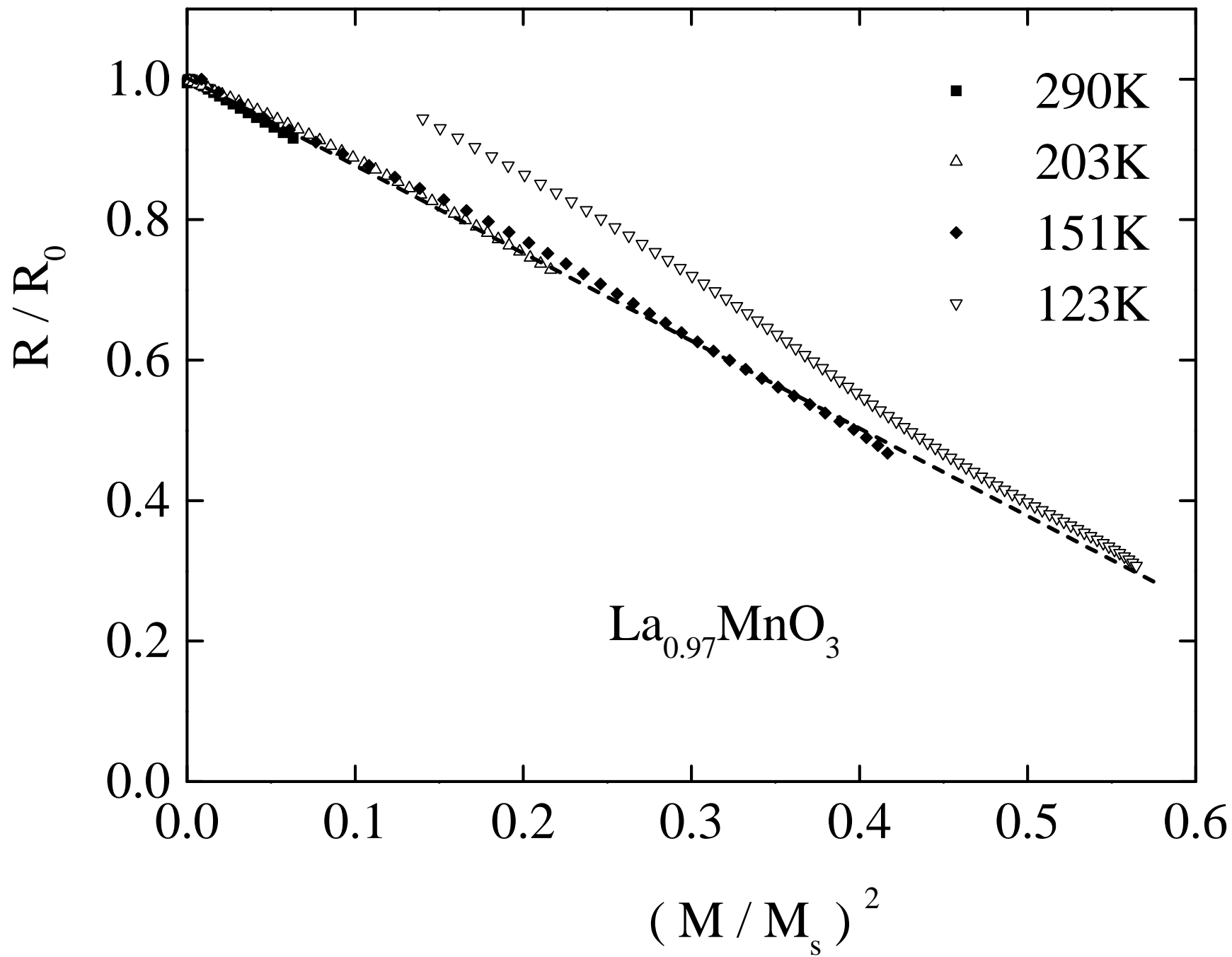
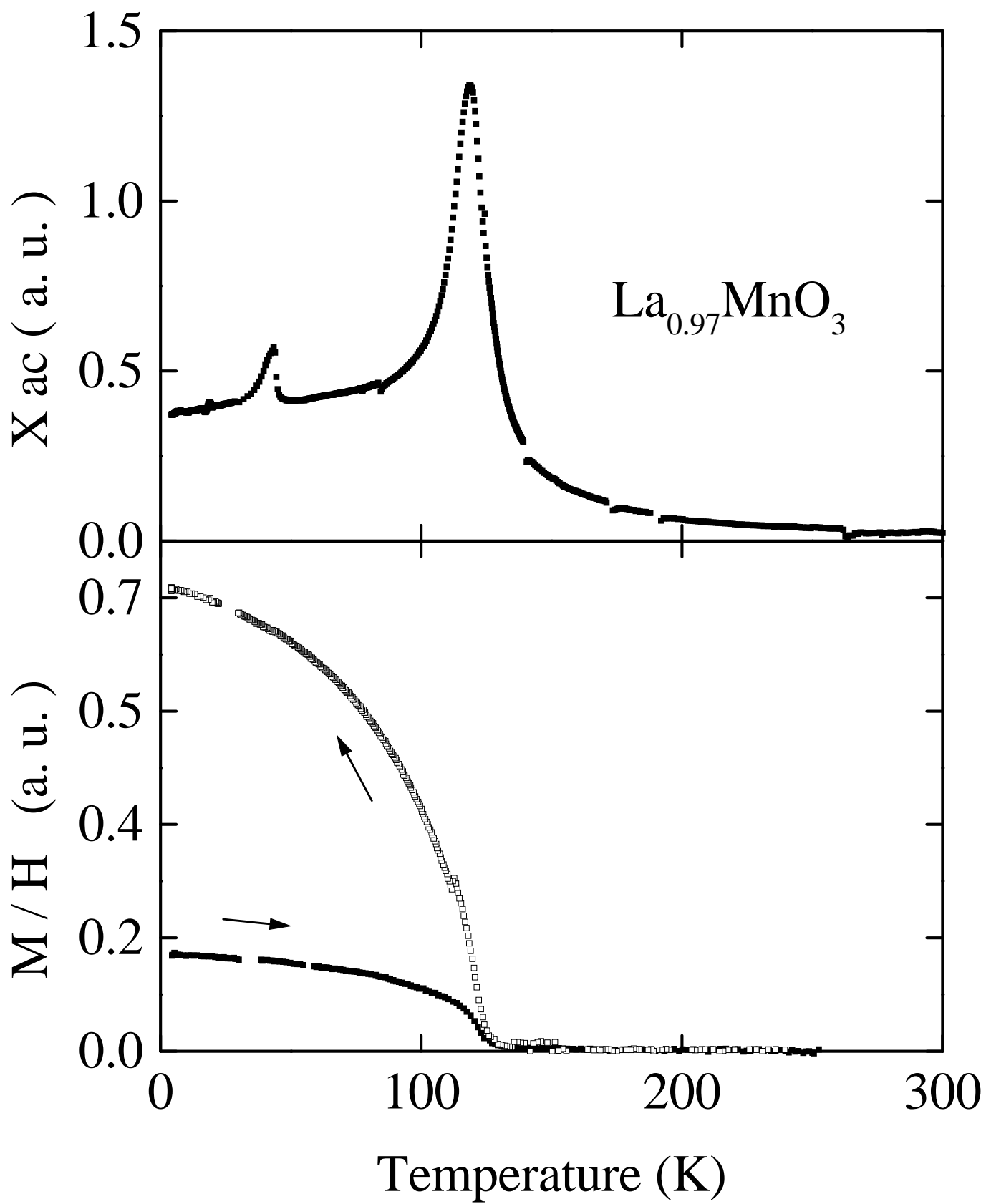
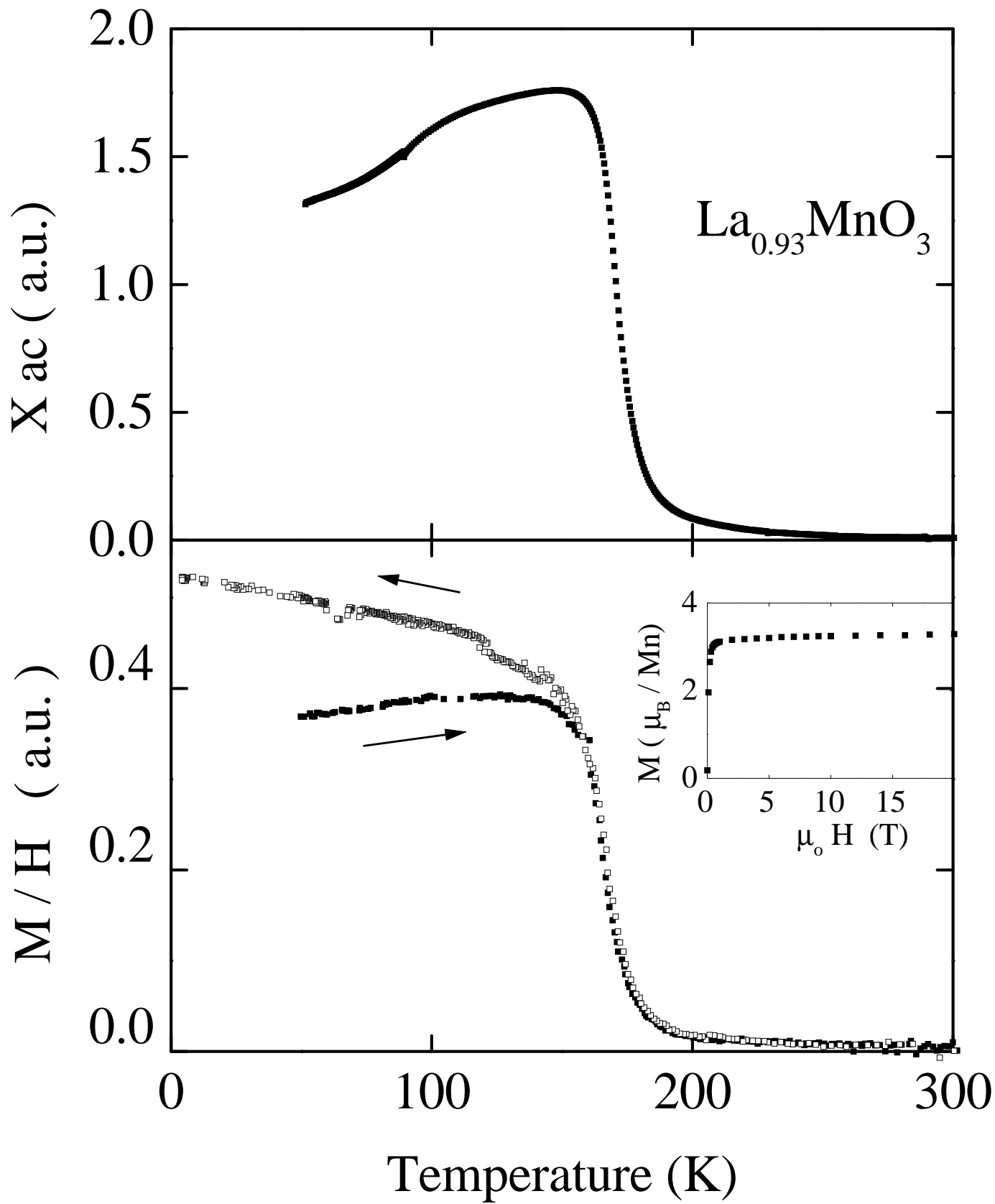
\end{document}